\documentclass[twocolumn,english,prb]{revtex4}
\usepackage[T1]{fontenc}
\usepackage[latin9]{inputenc}
\usepackage{color}
\usepackage{amsmath}
\usepackage{graphicx}
\usepackage{amssymb}
\usepackage{esint}

\makeatletter
\@ifundefined{textcolor}{}
{%
 \definecolor{BLACK}{gray}{0}
 \definecolor{WHITE}{gray}{1}
 \definecolor{RED}{rgb}{1,0,0}
 \definecolor{GREEN}{rgb}{0,1,0}
 \definecolor{BLUE}{rgb}{0,0,1}
 \definecolor{CYAN}{cmyk}{1,0,0,0}
 \definecolor{MAGENTA}{cmyk}{0,1,0,0}
 \definecolor{YELLOW}{cmyk}{0,0,1,0}
 }

\@ifundefined{definecolor}
 {\usepackage{color}}{}
\usepackage{babel}\usepackage[unicode=true, pdfusetitle,
 bookmarks=true,bookmarksnumbered=false,bookmarksopen=false,
 breaklinks=false,pdfborder={0 0 1},backref=false,colorlinks=false]{hyperref}\@ifundefined{definecolor}
 {\usepackage{color}}{}

\makeatletter\@ifundefined{textcolor}{}{\definecolor{BLACK}{gray}{0}\definecolor{WHITE}{gray}{1}\definecolor{RED}{rgb}{1,0,0}\definecolor{GREEN}{rgb}{0,1,0}\definecolor{BLUE}{rgb}{0,0,1}\definecolor{CYAN}{cmyk}{1,0,0,0}\definecolor{MAGENTA}{cmyk}{0,1,0,0}\definecolor{YELLOW}{cmyk}{0,0,1,0}}\@ifundefined{definecolor}{}{}\makeatother\makeatother

\makeatother

\usepackage{babel}

\begin{document}

\title{Spectral analysis for the iron-based superconductors: Anisotropic
spin fluctuations and fully gapped $s^{\pm}$-wave superconductivity}

\author{Junhua Zhang$^{1}$, Rastko Sknepnek$^{1,2}$, and Jörg Schmalian$^{1}$}

\affiliation{$^{1}$Department of Physics and Astronomy and Ames Laboratory, Iowa
State University, Ames, IA 50011, USA\\
 $^{2}$Department of Materials Science and Engineering Northwestern
University, Evanston, IL 60208, USA }

\date{\today}
\begin{abstract}
Spin fluctuations are considered to be one of the candidates that
drive a sign-reversed $s^{\pm}$ superconducting state in the iron
pnictides. In the magnetic scenario, whether the spin fluctuation
spectrum exhibits certain unique fine structures is an interesting
aspect for theoretical study in order to understand experimental observations.
We investigate the detailed momentum dependence of the short-range
spin fluctuations using a 2-orbital model in the self-consistent fluctuation
exchange approximation and find that a common feature of those fluctuations
that are capable of inducing a fully gapped $s^{\pm}$ state is the
momentum anisotropy with lengthened span along the direction transverse
to the antiferromagnetic momentum transfer. Performing a qualitative
analysis based on the orbital character and the deviation from perfect
nesting of the electronic structure for the 2-orbital and a more complete
5-orbital model, we gain the insight that this type of anisotropic
spin fluctuations favor superconductivity due to their enhancement
of intra-orbital, but inter-band, pair scattering processes. The momentum
anisotropy leads to elliptically shaped magnetic responses which have
been observed in inelastic neutron scattering measurements. Meanwhile,
our detailed study on the magnetic and the electronic spectrum shows
that the dispersion of the magnetic resonance mode in the nearly isotropic
$s^{\pm}$ superconducting state exhibits anisotropic propagating
behavior in an upward pattern and the coupling of the resonance mode
to fermions leads to a dip feature in the spectral function. 
\end{abstract}


\maketitle

\section{introduction}

Since the discovery of the iron-based superconductors,\cite{Kamihara08}
intensive research has been carried out and great progress has been
made to understand the microscopic mechanism of superconductivity
and the interplay between superconducting (SC), magnetic and structural
transitions. Still, many issues remain unclear such as the the pairing
mechanism, pairing symmetry, gap structure and the nature of the magnetism.
As the electron-phonon coupling seems to be inadequate to account
for the relatively high transition temperatures,\cite{Boeri08} electronic
interactions will likely play a role in the pair formation. Based
on the experience in high temperature cuprates, organic superconductors,
and heavy fermion superconductors with their commonality of proximity
to magnetism, comparative analyses have been made to show the similarities
and differences between different classes of superconductors.\cite{Scalapino10}
This directs the attention to the long discussed magnetic pairing
mechanism.\cite{Monthoux07nature} Renormalization group studies\cite{Chubukov08,FWang09}
on the iron pnictides show that the underlying electronic structure,
when close to perfect nesting, develops an unconventional pairing
interaction enabled by interband pair-hopping and further amplified
by antiferromagnetic correlations. Therefore electronic and magnetic
fluctuations, which are active in the vicinity of magnetic transition,
have been listed as one of the candidates to mediate Cooper pairs
in the material.\cite{Mazin08,Kuroki08,Mazin09,Chubukov08,FWang09,Ikeda10FLEX}
The pairing symmetry and gap structure are among the most interesting
topics as they contain most relevant information on pairing mechanism.
However there is no consensus on them when explored using different
experimental methods and in different members of the material. Further,
although being considered as layered structure, the material should
be considered anisotropic, but three dimensional. Some evidence shows
gap nodes along the $c$-axis.\cite{Tanatar10} Even for the in-plane
gap structure, some experiments support fully gapped structure and
some indicate in-plane gap nodes. Theoretical explorations suggest
the strongest pairing instability corresponds to the sign-reversed
$s^{\pm}$-wave state, without ruling out other forms like $d$-wave
and conventional $s$-wave. Even in the sign-reversed $s^{\pm}$ state,
depending on detailed Fermi surface (FS) configuration, different
form factors of the Cooper pair could be developed, resulting in either
fully gapped $s^{\pm}$-wave spectrum on all FS sheets or nodal $s^{\pm}$-wave
with sign change on the electron pockets.\cite{FWang10,Thomale10,Kemper10}
Besides, associated with the multiband character, evidence of multiple
gaps is found in some systems. It has been clear that the material
is metallic in the parent compound with well-established long range
antiferromagnetic (AFM) order. But with noticeable electronic correlations
the nature of magnetism is still under debate.

The magnetic excitation spectrum carries important information on
the nature of magnetism and the characteristics of superconductivity.
For the latter, it has been discussed in the context of cuprates that
an observation of a sharp quasiparticle-like resonance peak in the
spin fluctuation spectrum with the onset of superconductivity may
strongly indicate a sign change in the gap structure due to the superconducting
coherence factors. And the analogous discussion has been applied to
the iron pnictides\cite{Maier08,Korshunov08,Maier09,Seo09} based
on the random phase approximation and the mean-field BCS approximation,
showing that a strong spin resonance occurs in the $s^{\pm}$-wave
SC state. This indicates that the spin resonance phenomenon is compatible
with the $s^{\pm}$-wave SC gap. Meanwhile, as a momentum resolved
probe of the spin correlation, inelastic neutron scattering (INS)
experiments have reported the observation of resonance mode in both
hole-doped and electron-doped 122 system\cite{Christian08,Lumsden09,Chi09,Li09}
as well as in the $\mathrm{Fe}\mathrm{Te}_{1-x}\mathrm{Se}_{x}$ system.\cite{Lumsden10,Li10}
Notice that the spin resonance is a consequence of the sign-reversed
gap opening in the quasiparticle spectrum not an evidence for the
magnetic pairing glue. In order to reveal the relationship between
AFM fluctuations and superconductivity in the iron-based materials,
more detailed inspections on the structure of the spin fluctuations
are needed. Recently, INS measurements observe the same type of anisotropic
feature in the magnetic spectrum both in the normal and in the SC
state of the 122 system.\cite{Lester09,Diallo10,HF_Li10} This anisotropy
is characterized by larger broadening along the transverse direction
with respect to the AFM wave vector $\mathbf{Q}_{\text{AFM}}$ in
momentum space. Ref.\cite{HF_Li10} also shows no changes observed
in the spatial correlations through $T_{c}$, which is consistent
with the magnetic scenario in that the onset of superconductivity
does not change magnetic correlation length. Early theoretical exploration
on the short-range spin-fluctuation induced superconductivity has
argued that magnetic fluctuations throughout an extended momentum
region near AFM wave vector $\mathbf{Q}_{\text{AFM}}$ are relevant
to superconductivity. Thus it raises a question: is the observed anisotropic
feature of the spin fluctuations, i.e., larger broadening along the
transverse direction in momentum space, consistent with superconductivity
in this system?

On the other hand, as discussed for cuprates,\cite{Abanov99,Abanov01,Eschrig06review}
an important identification of the mediating boson, if it exists,
is from the fermionic spectrum which can be observed by angle-resolved
photoemission spectroscopy (ARPES). Indeed, ARPES has reported the
observations of kink feature in the electronic dispersion for the
hole-doped 122 system.\cite{Richard09,Wray08,Koitzsch09} However
there is discrepancy in its vanishing temperature among the observations
from different groups. If it is unique to the SC state, i.e., vanishing
above $T_{c}$, and the subtracted bosonic mode energy coincides with
the resonance energy, it would be a strong evidence for magnetic pairing
mechanism.

Motivated by these experimental facts, we perform a detailed investigation
of the spin and charge spectra in the normal and superconducting states,
in which the magnetic susceptibility and the SC gap function are determined
within the self-consistent fluctuation exchange (FLEX) approximation
using a 2-orbital model for iron pnictides. This itinerant model calculation
finds a fully gapped $s^{\pm}$-wave SC state driven by the enhanced
commensurate AFM correlation. Based on a systematic study on the momentum
structure of the short-range spin fluctuations, we find the same type
of anisotropy as that observed in INS measurements. To understand
the interplay between the fluctuation anisotropy and the $s^{\pm}$
superconductivity, we present a qualitative analysis through the orbital
contents and the deviation from perfect nesting of the electronic
structure for the 2-orbital and a more complete 5-orbital model. Meanwhile,
the calculated dispersion of the magnetic resonance mode exhibits
an anisotropic propagating pattern. And the calculated fermionic spectral
function shows the fingerprint of electron-mode coupling as observed
in ARPES.

\section{model}

We consider a Hamiltonian described by a 2-orbital tight-binding model
and on-site multiorbital electronic interactions,

\begin{align}
H & =\sum_{\mathbf{k},ab,\sigma}\varepsilon_{\mathbf{k}}^{ab}d_{\mathbf{k}a\sigma}^{\dagger}d_{\mathbf{k}b\sigma}+U\sum_{i,a}n_{ia\uparrow}n_{ia\downarrow}+U^{\prime}\sum_{i,a>b}n_{ia}n_{ib}\notag\\
 & +J_{H}\sum_{i,a>b,\sigma\sigma^{\prime}}d_{ia\sigma}^{\dagger}d_{ib\sigma^{\prime}}^{\dagger}d_{ia\sigma^{\prime}}d_{ib\sigma}\notag\\
 & +J^{\prime}\sum_{i,a\neq b}d_{ia\uparrow}^{\dagger}d_{ia\downarrow}^{\dagger}d_{ib\downarrow}d_{ib\uparrow}\label{eq:Hamiltonian}\end{align}
 where $n_{ia\sigma}=d_{ia\sigma}^{\dagger}d_{ia\sigma}$ is the occupation
number of the orbital $a$ with spin $\sigma$ at site $i$ and $n_{ia}=\sum_{\sigma}n_{ia\sigma}$
with the orbital index $a$($b$) standing for the Fe orbitals $d_{xz}$
and $d_{yz}$. The tight-binding description\cite{Raghu08} is given
by $\varepsilon_{\mathbf{k}}^{xy}=\varepsilon_{\mathbf{k}}^{yx}=-4t_{4}\sin k_{x}\sin k_{y}$
and $\varepsilon_{\mathbf{k}}^{aa}=-2t_{1}\cos k_{a}-2t_{2}\cos k_{b}-4t_{3}\cos k_{x}\cos k_{y}-\mu$
where $a,b=x(y)$ stand for $d_{xz}(d_{yz})$ orbitals and the momentum
components with $a\neq b$. We use the tight-binding parameters $t_{1}=-0.33$,
$t_{2}=0.385$, $t_{3}=-0.234$, and $t_{4}=-0.26$.\cite{Rastko09,Zhang09}
Near half filling this tight-binding model gives rise to the FS that
contains two hole pockets and two electron pockets for which we refer
to the hole pockets around $(0,0)$ and $(\pi,\pi)$ as $\alpha_{1}$
and $\alpha_{2}$ sheets, respectively, and the electron pockets around
$(\pi,0)$ and $(0,\pi)$ as $\beta_{1}$ and $\beta_{2}$ sheets,
respectively, in the unfolded (1 Fe per unit cell) Brillouin zone
(BZ). As noted by Kuroki \emph{et al}.\cite{Kuroki09} and Kemper
\emph{et al}.\cite{Kemper10}, the appearance of a hole pocket around
the $(\pi,\pi)$ point of the unfolded BZ is crucial to the formation
of fully gapped $s^{\pm}$ state. The predominant Fe-orbital character
distribution on each FS sheet has been analyzed in Ref.\cite{Graser09}.
In the 5-orbital tight-binding description,\cite{Kemper10} a third
orbital $d_{xy}$ predominantly contributes to the hole pocket at
the $(\pi,\pi)$ point and partially to the $(\pi,0)$ and $(0,\pi)$
electronic pockets. Although the 2-orbital model does not include
the third predominant orbital composition on the FS, as we will show
later, the intra-orbital nesting configuration of it remains similar
to that of the 5-orbital description, which is believed to play an
important role for the magnetic fluctuation and the superconductivity
in the itinerant picture. Thus this simplified 2-orbital description
qualitatively captures the key features of the electronic structure
near the Fermi energy, serving as a good starting point for the understanding
of the interplay between magnetism and superconductivity in the Fe-based
superconductors.

The on-site interactions consist of the intra- and inter-orbital Coulomb
repulsions controlled by the coupling constant $U$ and $U^{\prime}$,
the inter-orbital Hund's rule coupling $J_{H}$ and pair hopping term
$J^{\prime}$. For the bare Coulomb interaction, due to rotational
symmetry, the coupling constants are related by $U=U^{\prime}+2J$
and $J=J_{H}=J^{\prime}$. However, as discussed in the Ref.\cite{Zhang09},
when they are parameters in an approximate theory such as FLEX which
ignores vertex corrections, they are not identical to the bare Coulomb
matrix elements but should be considered as low energy coupling parameters
that have been renormalized by high energy excitations. Therefore
we study cases with various parameter values and present the typical
results here.

Using FLEX approximation\cite{Bickers89} for the SC state\cite{Bickers94,Scalapino94},
one self-consistently calculates the single-particle propagators renormalized
by interactions due to the exchange of spin and charge/orbital fluctuations
and the spin and charge/orbital correlation renormalized by the polarization
of dressed quasiparticles. In the SC state, Cooper-pair condensation
causes a finite anomalous self-energy, which can be treated as the
SC order parameter, and induces a contribution from the anomalous
quasiparticle propagator to the fluctuations. In this work we consider
only singlet pairing and assume time reversal symmetry in the system.

The complete multiorbital FLEX equations can be found in the Ref.\cite{Takimoto04}.
Here we present only expressions that are relevant to our discussions.
The magnetic susceptibility is theoretically calculated using the
Matsubara frequency as\begin{equation}
\chi_{s}\left(\mathbf{q},i\nu_{n}\right)=\sum_{aa,bb}\chi_{s}^{aa,bb}\left(\mathbf{q},i\nu_{n}\right),\label{eq:physical spin susceptibility}\end{equation}
 with bosonic Matsubara frequency $\nu_{n}=2n\pi T$ and the orbital-resolved
spin susceptibility calculated in the orbital-matrix expression\begin{eqnarray}
\hat{\chi}_{s}\left(\mathbf{q},i\nu_{n}\right) & = & \left[\hat{1}-\hat{\chi}_{s,0}\left(\mathbf{q},i\nu_{n}\right)\hat{U}_{s}\right]^{-1}\hat{\chi}_{s,0}\left(\mathbf{q},i\nu_{n}\right)\label{eq:matrix spin susceptibility}\end{eqnarray}
 where $\hat{U}_{s}$ is the spin-sector interaction matrix given
by $U_{s}^{aa,aa}=U$, as well as $U_{s}^{ab,ab}=U^{\prime}$, $U_{s}^{ab,ba}=J^{\prime}$,
and $U_{s}^{aa,bb}=J_{H}$ if $a\neq b$.\cite{Zhang09} Here the
bare susceptibility in the superconducting state is calculated as\begin{align}
 & \chi_{s,0}^{ab,a^{\prime}b^{\prime}}(\mathbf{q},i\nu_{n})=\notag\\
 & -\frac{T}{N}\sum_{\mathbf{k},\omega_{n}}\Bigl[G^{ba^{\prime}}(\mathbf{k+q},i\omega_{n}+i\nu_{n})G^{b^{\prime}a}(\mathbf{k},i\omega_{n})\notag\\
 & \ \ \ \ \ \ \ \ \ +F^{bb^{\prime}}(\mathbf{k+q},i\omega_{n}+i\nu_{n})F^{a^{\prime}a}(\mathbf{k},-i\omega_{n})\Bigl]\label{eq:bare spin susceptibility}\end{align}
 with the dressed normal and anomalous single-particle propagators
determined by solving the coupled Dyson-Gorkov's equations\begin{align}
\hat{G}(k) & =\hat{G}_{0}(k)+\hat{G}_{0}(k)\hat{\Sigma}(k)\hat{G}(k)+\hat{G}_{0}(k)\hat{\Phi}(k)\hat{F}(k^{\prime})^{*},\notag\\
\hat{F}(k) & =\hat{G}_{0}(k)\hat{\Sigma}(k)\hat{F}(k)-\hat{G}_{0}(k)\hat{\Phi}(k)\hat{G}(k^{\prime})^{*},\label{eq:Dyson-Gorkov equation}\end{align}
 where $k=(\mathbf{k},i\omega_{n})$, $k^{\prime}=(-\mathbf{k},i\omega_{n})$
with fermionic Mastubara frequency $\omega_{n}=(2n+1)\pi T$. The
normal and anomalous self-energies \begin{align}
 & \Sigma^{ab}(\mathbf{k},i\omega_{n})\notag\\
 & =\frac{T}{N}\sum_{\mathbf{q},\nu_{n}}\sum_{m,n}\Gamma_{N}^{am,bn}(\mathbf{q},i\nu_{n})G^{mn}(\mathbf{k-q},i\omega_{n}-i\nu_{n}),\label{eq:normal self-energy}\\
 & \Phi^{ab}(\mathbf{k},i\omega_{n})\notag\\
 & =\frac{T}{N}\sum_{\mathbf{q},\nu_{n}}\sum_{m,n}\Gamma_{A}^{am,nb}(\mathbf{q},i\nu_{n})F^{mn}(\mathbf{k-q},i\omega_{n}-i\nu_{n}),\label{eq:anomalous self-energy}\end{align}
 are related to the renormalized spin and charge/orbital susceptibility
through the normal and anomalous interaction vertices $\hat{\Gamma}_{N}$
and $\hat{\Gamma}_{A}$.

Carrying out the analytical continuation from Mastubara frequencies
to the real frequencies numerically using Pad\textcolor{black}{é approximant,
the dynamical spin susceptibility is given by\begin{equation}
\chi_{s}(\mathbf{q},\omega)=\chi_{s}\left(\mathbf{q},i\nu_{n}\rightarrow\omega+i\delta\right),\label{eq:dynamical spin susceptibility}\end{equation}
 whose imaginary part $\mathrm{Im}\chi_{s}$ directly relates to INS
intensity. Simultaneously the quasiparticle spectral function is obtained
as}\begin{equation}
A\left(\mathbf{k},\omega\right)=-\frac{1}{\pi}\mathrm{Im}\left[\sum_{a}G^{aa}\left(\mathbf{k},i\omega_{n}\rightarrow\omega+i\delta\right)\right],\label{eq:spectral function}\end{equation}
 which corresponds to ARPES intensity.

\section{superconducting gap structure}

Within the framework of the self-consistent fluctuation exchange approximation,
we search for stable SC solution using the above itinerant model at
different coupling constants, doping levels and temperatures. Further
we examine the features of the magnetic response for systems that
develop superconductivity induced by exchange of short-range fluctuations.
In FLEX formalism both spin and charge/orbital fluctuations contribute
to the pairing interaction, but the major contribution to the pairing
glue comes from spin fluctuations for the parameter regime we are
studying. The nested structure of the FS leads to peak in the magnetic
susceptibility near the AFM wave vector $\mathbf{Q}_{\text{AFM}}=(0,\pi),\ (\pi,0)$,
which is strongest at half-filling, $n=2.0$ per site, indicating
its correspondence to the parent compound.

The calculations are performed on imaginary frequency axis for a lattice
of $64\times64$ sites with $8192$ Matsubara frequencies. When self-consistently
solving the FLEX equations, the convergence of iterations is considered
achieved when the maximum relative difference between two consecutive
iterations of the self-energy element, $\Sigma^{ab}(\mathbf{k},i\omega_{n})$
or $\Phi^{ab}(\mathbf{k},i\omega_{n})$, is less than $10^{-6}$.
\begin{figure}
\includegraphics{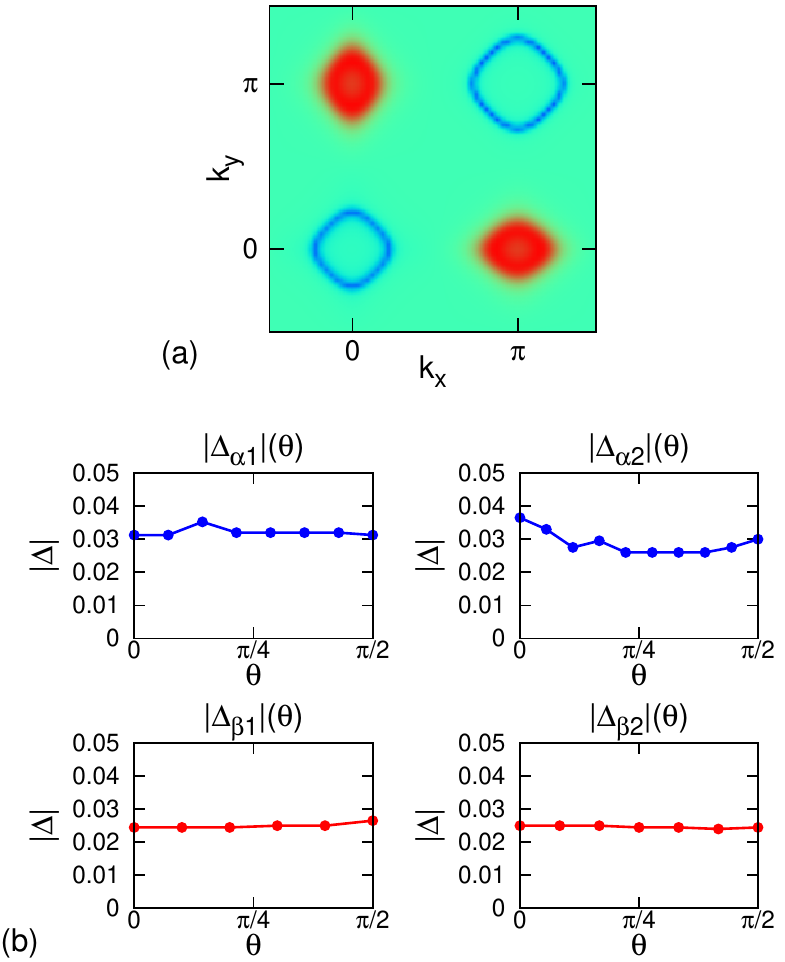} \caption{(Color online) (a) The renormalized Fermi surface at $T\approx0.1T_{c}$
for $n=1.88$. (b) Gap magnitude $|\Delta|$ versus angle around each
FS sheet. There is an overall sign change in $\Delta$ between the
$\alpha$ and $\beta$ sheets.\label{fig:Fermi-surface-configuration} }

\end{figure}

For the current model we only find stable SC solutions in the hole
doped regime, with the particle density $1.85\le n\le1.90$ per site,
driven by short-range spin fluctuations for certain range of coupling
constants. The achieved SC states are of $s^{\pm}$-wave symmetry:
fully gapped on each FS sheet with an overall sign change between
the $\alpha$ and $\beta$ sheets. To illustrate the momentum structure
of the gap function, we show the results for a typical set of coupling
constants $U=1.5,\ U^{\prime}=1.2$ and $J=J^{\prime}=0.8$ with particle
density $n=1.88$ at $T=0.001$. In this case the system exhibits
a transition from the paramagnetic normal state to the SC state at
$T_{c}=0.0075$ and the static spin susceptibility $\chi_{s}(\mathbf{q},\omega=0)$
shows well pronounced peaks at the commensurate wave vector $\mathbf{Q}_{\text{AFM}}=(0,\pi),\ (\pi,0)$
both in the normal and in the SC  state. Note that we do not assign
specific units to the parameters and quantities: The coupling constants
scale with the hopping parameters and all energies scale with the
SC gap magnitude while the temperatures scale with $T_{c}$. In Fig.\,\ref{fig:Fermi-surface-configuration}
(a), we show the renormalized FS configuration obtained from the intensity
projection of the Cooper-pair wave function $F(\mathbf{k},\omega)$.
Taking the small damping parameter $\delta=0.002$ in the analytical
continuation for the $T=0.001$ solution and subtracting the gap magnitude
from the spectral function $A(\omega)$ at the Fermi wave vectors
$\mathbf{k}_{F}$ for each FS sheet, the variation in the gap magnitude
on each separated FS sheet is plotted in \textcolor{black}{Fig.\,\ref{fig:Fermi-surface-configuration}
(b). Clearly in this case the gap is nearly isotropic on each pocket.}
The ratio $2\Delta/T_{c}\sim6-8$ implies a strong coupling system.

\section{short-range spin fluctuations}

In this section we investigate the momentum structure of the short-range
magnetic fluctuations that mediate superconductivity and make connection
with the magnetic response measured in INS experiments. Our main points
are as follows:
\begin{itemize}
\item The short-range spin fluctuations that are capable of driving the
fully gapped $s^{\pm}$ superconductivity generally exhibit an anisotropy
in momentum space with $\mathbf{q}$-width larger along the direction
transverse to $\mathbf{Q}_{\text{AFM}}$ than along the longitudinal
direction. This can be understood by examining the intra-orbital scattering
processes in systems away from perfect nesting.
\item The momentum structure of the spin excitations exhibits the same type
of anisotropy, which, in the SC state, gives rise to an elliptical
shape of the spin resonance mode. Further, the resonance mode disperses
with increasing energy in the pattern broadening more rapidly along
the transverse than along the longitudinal direction. This anisotropic
dispersion of the resonance mode associated with the intrinsic anisotropy
of the mode leads to more elliptically shaped $\mathbf{q}$-image.
\item The dispersion of the magnetic resonance shows an upward pattern with
increasing energy in the nearly isotropic $s^{\pm}$ state with commensurate
magnetic peak. But the weight of the mode decays dramatically and
vanishes above the particle-hole threshold.
\item In the strong coupling approach, the resonance energy is affected
by the SC gap magnitude and the magnetic correlation strength. 
\end{itemize}
To illustrate these points, we present the results for the magnetic
susceptibility $\chi_{s}\left(\mathbf{q},\omega\right)$ calculated
using the typical set of parameters mentioned in the previous section
followed with discussions. As Park \emph{et al.} \cite{Hinkov10}
show that the unfolded BZ description of the magnetic spectrum in
the paramagnetic state is justified, our spin-fluctuation spectrum
calculated in the BZ with 1 Fe ion is discussed below. The qualitative
agreement of the calculated anisotropy with that observed in INS,
in turn, suggests that the magnetic spectrum originates predominantly
from the Fe-sublattice. In the following we refer to the transverse
(TR) or longitudinal (LO) direction as the direction transverse or
longitudinal to the corresponding AFM momentum transfer $\mathbf{Q}_{\text{AFM}}$.

\subsection*{Results}

\begin{figure}
\includegraphics{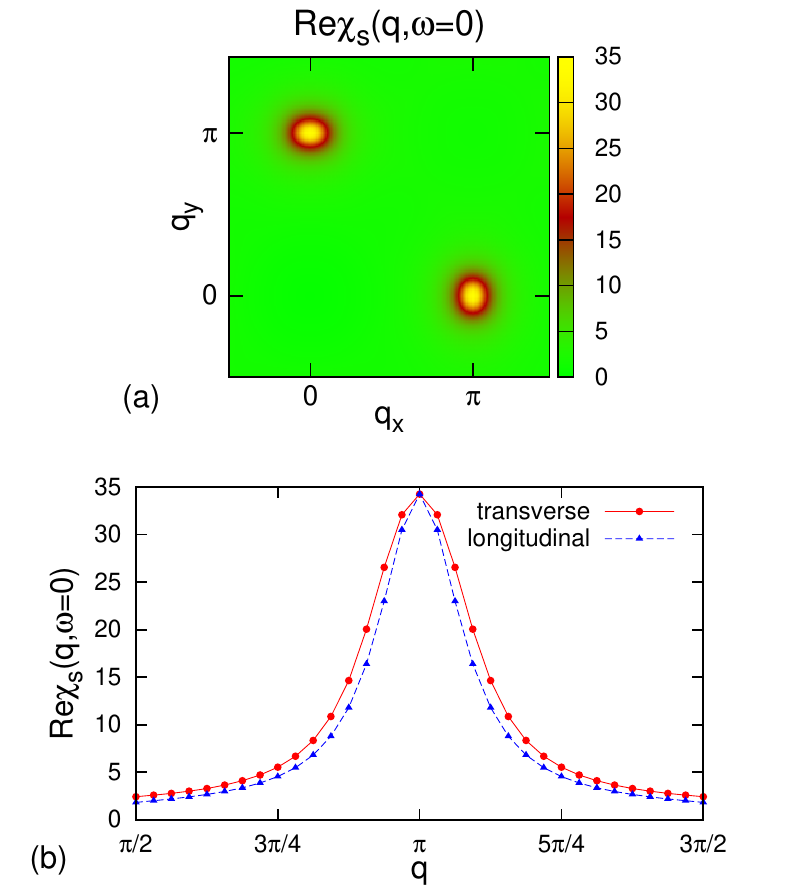}

\textcolor{black}{\caption{\label{fig:static}(Color online) The $\mathbf{q}$-anisotropy of
the static spin susceptibility. (a) is the intensity plot of $\mathrm{Re}\chi_{s}(\mathbf{q},\omega=0)$
in the momentum space and (b) shows the scans along the TR (red solid
circle) and LO (blue solid triangle) direction, respectively.}
}

\end{figure}

\begin{figure}
\includegraphics{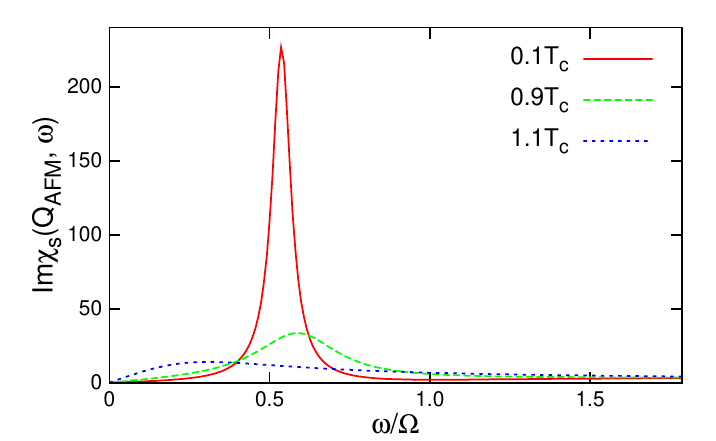} \caption{(Color online) Magnetic spectrum at the momentum transfer $\mathbf{Q}_{\mathrm{AFM}}$
slightly above ($T\approx1.1T_{c}$) and below ($T\approx0.9T_{c}$)
the transition temperature $T_{c}$, as well as deep in the superconducting
state ($T\approx0.1T_{c}$).\label{fig:resonance_T} }

\end{figure}

To analyze the momentum structure of the spin fluctuations, we begin
with the static spin susceptibility $\mathrm{Re}\chi_{s}(\mathbf{q},\omega=0)$.
As shown in Fig.\,\ref{fig:static}\textcolor{black}{, besides that
the static response achieves strongest enhancement at }$\mathbf{Q}_{\text{AFM}}$\textcolor{black}{,
spins are correlated spatially in an anisotropic manner with the largest
span along the TR direction} and smallest along the LO direction in
momentum space. This results in an elliptically shaped momentum structure.
Our systematic study shows that the degree of anisotropy increases
with the deviation from the perfect nesting in the electronic structure.
Moreover, the calculation for the temperature right above and right
below the transition temperature shows that the spatial correlation
does not change through $T_{c}$, reflecting the fact that spin-fluctuation
induced superconductivity does not modify magnetic correlation length
when entering SC phase, although it is the same electrons that contribute
to the magnetic and SC properties. This is in agreement with the INS
observations.\cite{HF_Li10}

\begin{figure}
\includegraphics{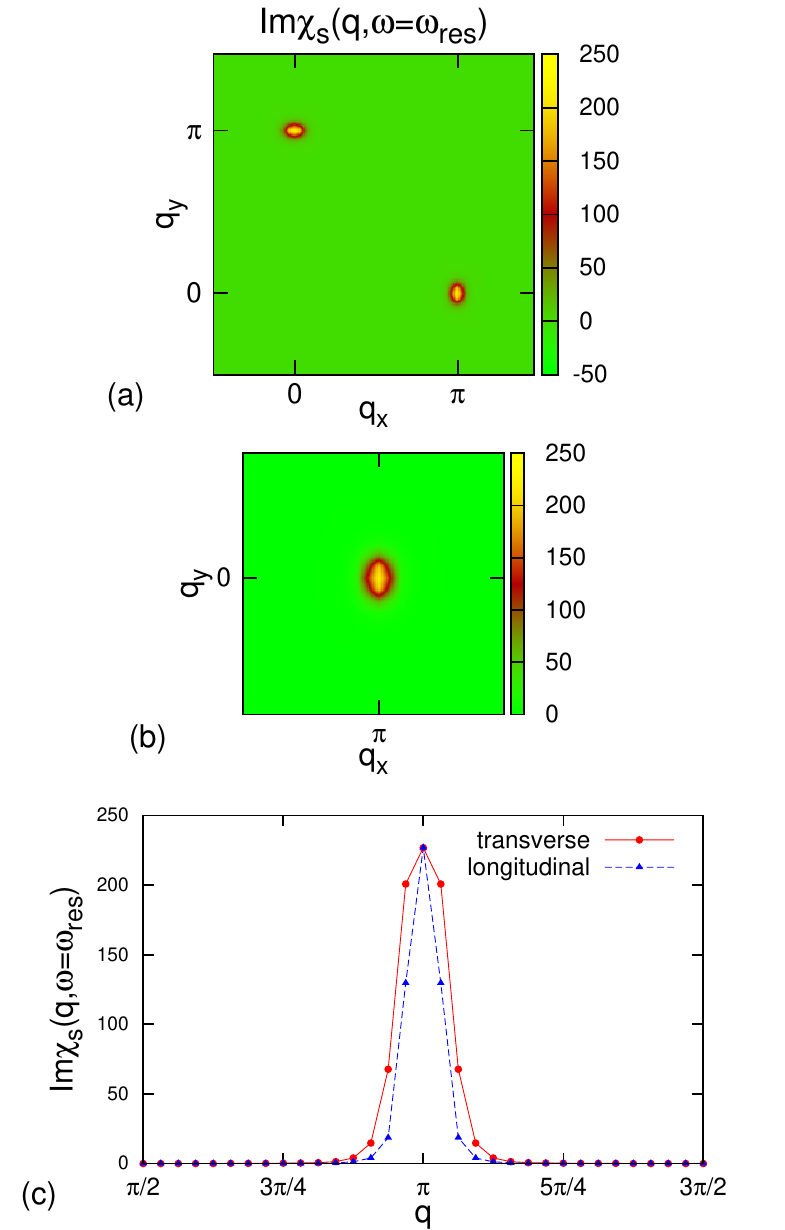}

\textcolor{black}{\caption{\label{fig:anisotropic_resonance}(Color online) The $\mathbf{q}$-anisotropy
of the spin resonance mode. (a)  is the intensity plot of $\mathrm{Im}\chi_{s}(\mathbf{q},\omega_{\text{res}})$
at the resonance energy in the momentum space and (b) gives a zoom-in
image of the resonance mode at $(\pi,0).$ (c) shows the scans along
the TR (red solid circle) and LO (blue solid triangle) direction,
respectively.}
}

\end{figure}

\begin{figure*}
\includegraphics{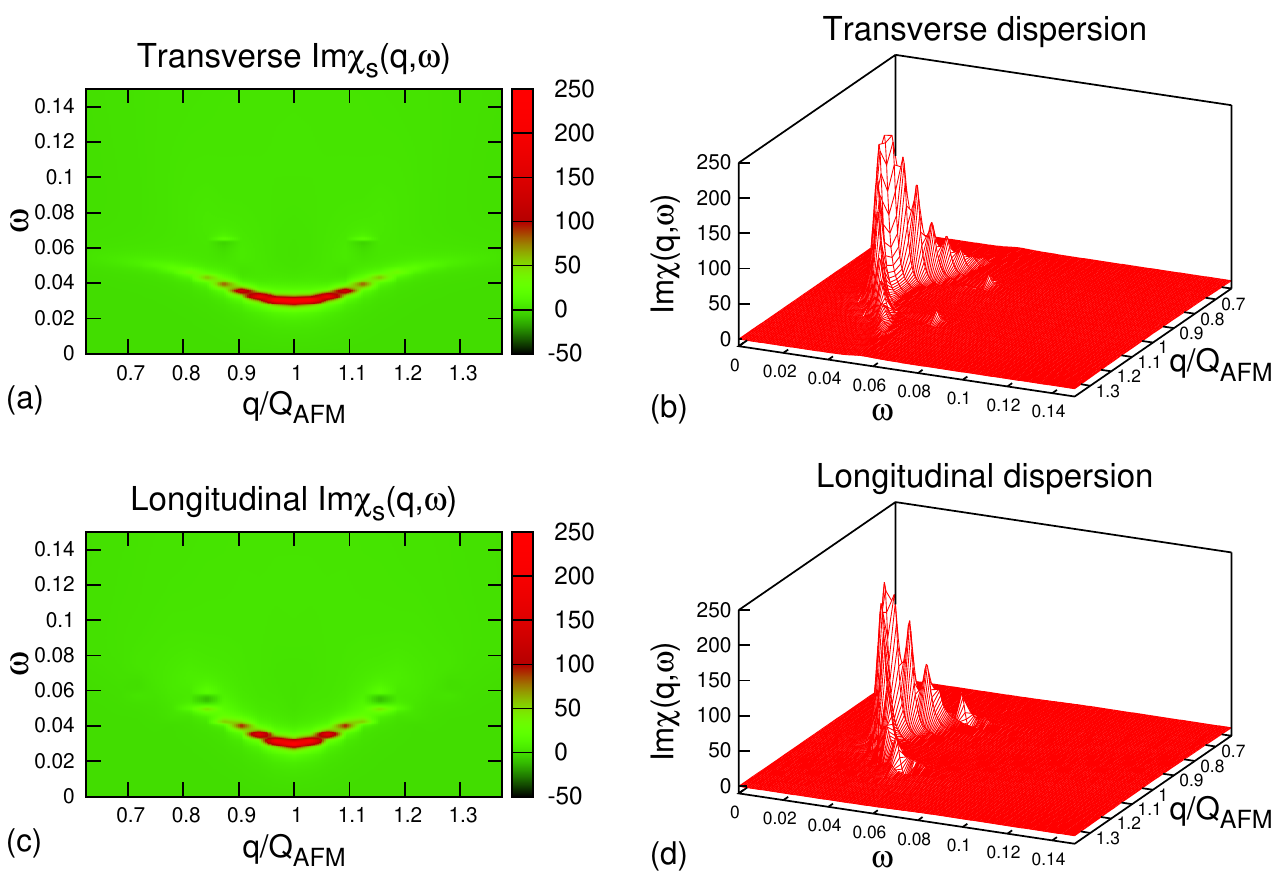}

\textcolor{black}{\caption{\label{fig:The-anisotropic-dispersive}(Color online) The anisotropic
dispersive behavior of the resonance mode along the TR and LO directions
with respect to $\mathbf{Q}_{\mathrm{AFM}}$. (a) and (c) are the
intensity plot along the TR and LO direction, respectively, while
(b) and (d) show the weight decay of the propagating mode.}
}

\end{figure*}

The imaginary part of the spin susceptibility $\mathrm{Im}\chi_{s}(\mathbf{q},\omega)$
contains information on the magnetic excitations. As discussed for
a sign-reversed SC gap structure, the most striking feature of the
magnetic spectrum is the appearance of a resonance mode at the characteristic
momentum transfer $\mathbf{Q}_{\text{AFM}}$ when entering SC phase
in spite of no long-range magnetic order. This sharp mode is of spin-excitonic
type in our model, originated from the Stoner enhancement factor $\left[\det\left\vert \hat{1}-\hat{\chi}_{s,0}\left(\mathbf{q},i\nu_{n}\right)\hat{U}_{s}\right\vert \right]^{-1}$.\cite{Abanov99,Eschrig06review}
Indeed, our calculation indicates a well-pronounced quasiparticle-like
peak as shown in Fig.\,\ref{fig:resonance_T}.\textcolor{black}{{}
In this figure, the results for the }spin susceptibility $\mathrm{Im}\chi_{s}(\mathbf{Q}_{\text{AFM}},\omega)$
as a function of frequency at the momentum transfer $\mathbf{\mathbf{Q}_{\text{AFM}}}$
for temperatures $T\approx1.1T_{c}$, $T\approx0.9T_{c}$ and $T\approx0.1T_{c}$
are presented. In the normal state the magnetic spectrum exhibits
a broad continuum associated with the overdamping feature of the spin
fluctuations. The transition to SC state modifies the spectrum by
pushing the spectral weight to higher energy and rapidly developing
a resonance mode as the temperature decreasing. The fact that this
mode is made out of a particle-hole bound state in the excitonic form
leads to an energy threshold taking the minimal value of sums of two
gap magnitudes on any pair of FS points connected by the momentum
transfer $\mathbf{\mathbf{Q}_{\text{AFM}}}$. Here we refer to this
threshold as $\Omega\equiv\min_{\mathbf{k}}\left(|\Delta_{\mathbf{k}}|+|\Delta_{\mathbf{k+Q}_{\text{AFM}}}|\right)$.
For the parameter set under discussion, the ratio of the resonance
energy to the threshold and to the SC transition temperature are roughly
$\omega_{\text{res}}/\Omega\approx0.6$ and $\omega_{\text{res}}/k_{B}T_{c}\approx4$,
which agrees with the experimental values measured for $\mathrm{K}$-
and $\mathrm{Co}$-doped $\mathrm{BaFe_{2}As_{2}}$.\cite{Christian08,Lumsden09}

Next we analyze the momentum dependence of the magnetic spectrum at
the resonance frequency $\omega_{\text{res}}$. Figure \ref{fig:anisotropic_resonance}\textcolor{black}{{}
shows the results for }$\mathrm{Im}\chi_{s}(\mathbf{q},\omega_{\text{res}})$
where a zoom-in $\mathbf{q}$-image of the mode at $(\pi,0)$ is given
in the middle. Similar to the static magnetic response, the $\mathbf{q}$-shape
of the resonance mode is also elliptical with maximal broadening along
the TR direction and minimal along the LO direction. This is an intrinsic
anisotropy of the magnetic spectrum not only in the SC state but also
existing in the normal state, which has been observed in the INS measurements.\cite{Lester09,HF_Li10}

\begin{figure}
\includegraphics{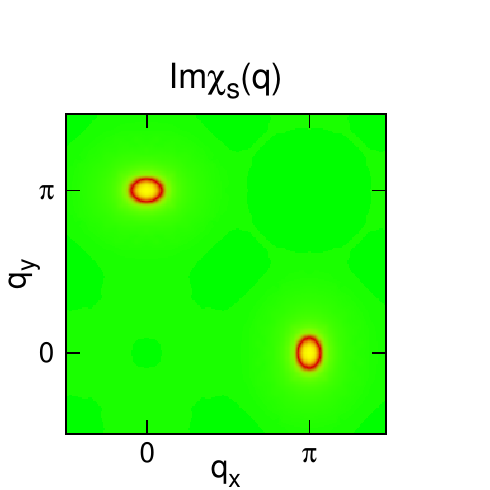}

\textcolor{black}{\caption{\label{fig:average}(Color online) The dynamical spin susceptibility
averaged over a small frequency window through the resonance energy:
$\frac{1}{2\Delta\omega}\int_{\omega_{\mathrm{res}}-\Delta\omega}^{\omega_{\mathrm{res}}+\Delta\omega}\mathrm{d}\omega\mathrm{Im}\chi_{s}(\mathbf{q},\omega)$
with \textcolor{black}{$\Delta\omega=\omega_{\text{res}}/4$} here. }
}
\end{figure}

\begin{figure*}
\includegraphics{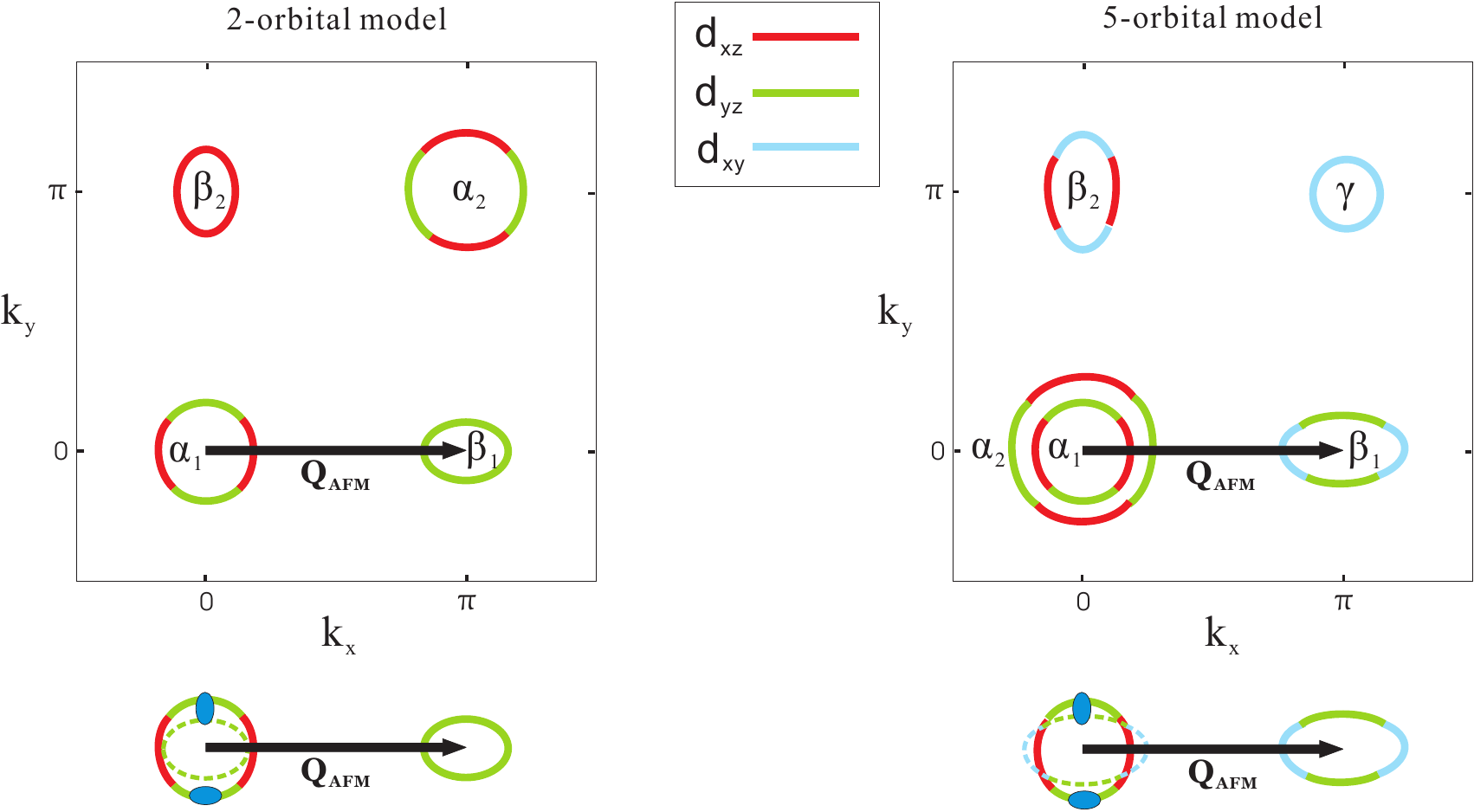} \caption{(Color online) Schematic illustration of the intra-orbital pair scattering
processes with momentum transfer $\mathbf{Q}_{\text{AFM}}$ for the
2-orbital and a more complete 5-orbital model. In the lower part,
by translating $\mathbf{Q}_{\mathrm{AFM}}$, the $\beta_{1}$-pocket
is moved to the position of the $\alpha_{1}$-pocket. For intra-orbital,
but inter-band, scattering to happen, the effective scattering vertices,
depicted by the small dark-blue ellipses, should be able to cover
the same orbital pieces on the two deviated FS sheets, \emph{i.e.},
the $\alpha_{1}$ sheet and the shifted $\beta_{1}$ sheet (dashed
line). One can perform similar operation for the $\alpha_{1}(\alpha_{2})$
and $\beta_{1}(\beta_{2})$ sheets in the 2-orbital model and for
the $\alpha_{1}(\gamma)$ and $\beta_{1}(\beta_{2})$ sheets in the
5-orbital model. Clearly, the transversely lengthened vertices are
more capable of inducing the intra-orbital, but inter-band, pair scattering
processes.\label{fig:Schematical-illustration} }

\end{figure*}

More interestingly, the propagation of the resonance mode also exhibits
an anisotropic behavior. \textcolor{black}{Figure \ref{fig:The-anisotropic-dispersive}
shows the dispersion of the resonance mode along the TR and LO directions}
at $T\approx0.1T_{c}$, deep in the SC state. Two features are associated
with the propagating behavior of the quasiparticle-like magnetic excitations
in a fully gapped nearly isotropic $s^{\pm}$ state driven by commensurate
short-range spin fluctuations: First, the resonance mode disperses
with increasing energy in an anisotropic pattern broadening most rapidly
along the TR rather than the LO direction as clearly seen in Fig.\textcolor{black}{\,\ref{fig:The-anisotropic-dispersive}}
(a) and (c), reminiscent of the anisotropic propagation of spin waves
in the spin density wave state of the 122 parent compound;\cite{Matan09}
second, the resonance mode disperses upwards in energy with dramatically
decreasing weight and vanishes above the particle-hole threshold $\Omega$
as shown in Fig.\textcolor{black}{\,\ref{fig:The-anisotropic-dispersive}}
(b) and (d). This upward dispersion is in contrast to the downward
pattern in the $d$-wave cuprate.\cite{Eremin05}

The anisotropy of the dispersion relation enhances the image ellipticity
of the measured magnetic response, if we consider a frequency average
of the spectrum over a small window through the resonance energy,
\emph{i.e.}, $\frac{1}{2\Delta\omega}\int_{\omega_{\mathrm{res}}-\Delta\omega}^{\omega_{\mathrm{res}}+\Delta\omega}\mathrm{d}\omega\mathrm{Im}\chi_{s}(\mathbf{q},\omega)$,
mimic the observation in INS.\textcolor{black}{{} This enhanced ellipticity
due to the combination of the intrinsic and dispersing anisotropy
of the resonance mode is shown in }Fig.\,\ref{fig:average}\textcolor{black}{,
where we take $\Delta\omega=\omega_{\text{res}}/4$.}

\subsection*{Discussion of the spin-fluctuation anisotropy and the $s^{\pm}$
superconductivity}

A systematic study of the 2- and a 3-orbital\cite{Daghofer10} models,
both in hole and in electron doped regions for a variety of coupling
constants, draws our attention to the connection between the momentum
anisotropy of the magnetic fluctuations and the $s^{\pm}$ superconductivity.
Our normal-state calculations show that different anisotropic pattern
of the fluctuations occurs at different parameter sets in different
model systems, either transversely or longitudinally lengthened. But
the development of short-range spin fluctuations centered at $\mathbf{Q}_{\text{AFM}}$
does not necessarily lead to superconductivity. For the various systems
we have studied, the establishment of a stable $s^{\pm}$ state is
generally associated with the transversely lengthened fluctuations.
This is the characteristic momentum structure of the static correlations
and of the magnetic spectra both in the normal and in the SC state.
It poses a question: is superconductivity sensitive to the momentum
structure of the magnetic glue?

Here we discuss how the specific momentum structure of the spin fluctuations
affects the $s^{\pm}$ superconductivity in the magnetic scenario
where the same electrons contribute to both the magnetic and the SC
properties. We gain the insight by recognizing the important role
played by the orbital weight on the FS sheets. As pointed out by Kemper
\emph{et al}.,\cite{Kemper10} the dominant pairing processes involve
intra-orbital scattering. The intra-orbital effective pairing interaction
vertex $\Gamma_{A}^{cc,cc}$ is dominated by the processes of exchanging
spin-1 fluctuations as \[
\Gamma_{A}^{cc,cc}\left(\mathbf{k},\mathbf{k}^{\prime},i\nu_{n}\right)\sim\frac{3}{2}\sum_{aa,bb}U_{s}^{cc,aa}\chi_{s}^{aa,bb}\left(\mathbf{k}-\mathbf{k}^{\prime},i\nu_{n}\right)U_{s}^{bb,cc}\]
 where $a,b,c$ are orbital indices and $\Gamma_{A}^{cc,cc}$ becomes
significant when the momentum transfer $\mathbf{k}-\mathbf{k}^{\prime}\sim\mathbf{Q}_{\text{AFM}}$.
It gives rise to the intra-orbital Cooper-pair formation through\[
\Phi^{cc}(\mathbf{k},i\omega_{n})=\frac{T}{N}\sum_{\mathbf{k}',\omega_{n}'}\Gamma_{A}^{cc,cc}(\mathbf{k},\mathbf{k}',i\omega_{n}-i\omega_{n}')F^{cc}(\mathbf{k}',i\omega_{n}')\]
which scatters a pair of $c$-orbital electrons on $\alpha(\beta)$
sheets to a pair of $c$-orbital electrons on $\beta(\alpha)$ sheets,
i.e., an inter-band scattering.

A schematic demonstration of the FS configuration, with predominant
iron $d$-orbital distribution indicated, for the current 2-orbital
model and for a more complete 5-orbital model are shown in Fig.\,\ref{fig:Schematical-illustration},
which were analyzed by Graser \emph{et al}.\cite{Graser09} and Kemper
\emph{et al}.\cite{Kemper10}. To illustrate the pair scattering between
two bands with the typical momentum transfer, $\beta_{1}$ pocket
is translated by $\mathbf{Q}_{\text{AFM}}$ to overlap with $\alpha_{1}$
pocket. As the electronic structure is away from perfect nesting,
the intra-orbital scattering is more supported by the transversely
lengthened pairing vertices than the longitudinally lengthened ones
which are depicted by the small dark-blue ellipses in the figure.
One can perform the same translation for other pairs of pockets separated
by $\mathbf{Q}_{\text{AFM}}$ in both models and draw the same conclusion.
Therefore, due to the orbital character of the microscopic electronic
structure and the deviation from perfect nesting, this type of anisotropic
momentum structure of short-range spin fluctuations favor the formation
of $s^{\pm}$ SC state, since the intra-orbital pairing processes
are made more plausible driven by transversely lengthened fluctuations.

\subsection*{Factors affecting the resonance energy}

In the strong coupling approach, the factors affecting the resonance
mode energy involve the SC gap magnitude $|\Delta|$ and the magnetic
correlation length $\xi$.\cite{Abanov99} Our systematic study indicates
that the resonance energy $\omega_{\text{res}}$ increases with increasing
gap magnitude but decreases with increasing correlation length.

\section{fermionic spectrum}

\begin{figure}
\includegraphics{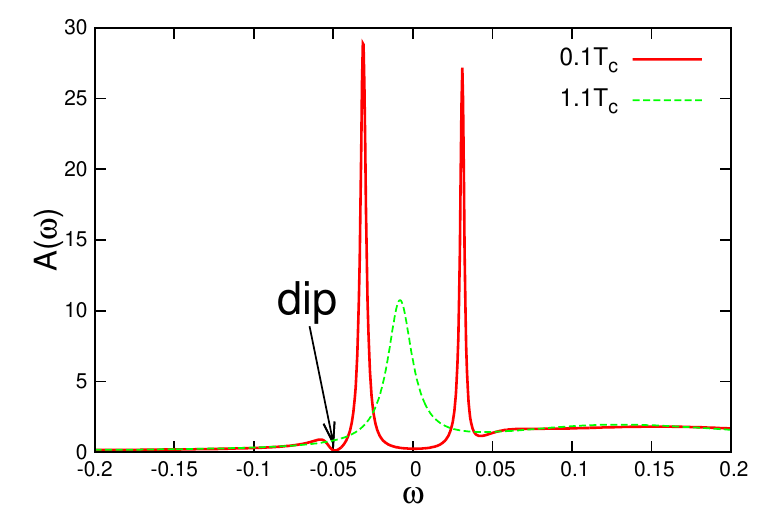}

\textcolor{black}{\caption{\label{fig:The-peak-dip-hump-feature}(Color online) The fermionic
spectral function $A(\omega)$ at a \textcolor{black}{$\mathbf{k}$-point
on the $\alpha_{1}$ sheet in the SC ($T\approx0.1T_{c}$) state (red
solid line) as well as in the normal ($T\approx1.1T_{c}$) state (green
dashed line). Notice} the peak-dip-hump feature in the negative frequency
regime when the system becomes superconducting. Comparing the SC state
data with the normal state data, we see that the spectral weight is
conserved.}
}

\end{figure}

As discussed for cuprates,\cite{Abanov99,Abanov01,Eschrig06review}
the impact of the mediating bosonic modes on fermions leaves fingerprint
in the fermionic spectrum, the SC spectral function, if the bosonic
excitations are gapped quasiparticles such as optical phonons in the
conventional superconductors. This gives rise to the kink feature
in the electronic energy dispersion observed in ARPES measurements.
The signature of electron-mode coupling is believed to be linked to
the pairing. If short-range spin fluctuations mediate Cooper pairs,
with the emergence of the gapped quasiparticle-like spin resonance
mode in the superconducting state, large fermionic decay occurs by
exchange of the magnetic resonance mode in scattering processes. This
leaves a dip in the electronic spectral function at the energy $\omega_{\mathrm{dip}}=\left|\Delta_{\mathbf{k}+\mathbf{Q}_{\text{AFM}}}\right|+\omega_{\mathrm{res}}$
where $\left|\Delta_{\mathbf{k}+\mathbf{Q}_{\text{AFM}}}\right|$
indicates the SC gap magnitude at the Fermi point connected by the
AFM wave vector. Indeed, our calculation does show the dip feature
in the electronic spectral function. \textcolor{black}{As shown in
Fig.\,\ref{fig:The-peak-dip-hump-feature}, the spectral function
at a $\mathbf{k}$-point on the $\alpha_{1}$ sheet exhibits the characteristic
peak-dip-hump behavior in $A_{\mathbf{k}}(\omega)$ with the dip position
$\omega_{\text{dip}}\approx0.05$ to be the sum of the resonance energy
$\omega_{\text{res}}\approx0.03$ and the gap magnitude $\left|\Delta_{\mathbf{k}+\mathbf{Q}_{\text{AFM}}}\right|\approx0.02$
at the point on the other band connected by $\mathbf{Q}_{\text{AFM}}$.
The normal state data at the same $\mathbf{k}$-point is also plotted
in Fig.\,\ref{fig:The-peak-dip-hump-feature} to show the conservation
of spectral weight.}

\section{conclusions}

In summary, we have investigated the momentum structure of the short-range
magnetic fluctuations that drive the nearly isotropic $s^{\pm}$ superconductivity
using a microscopic model for the Fe-based superconductors in the
self-consistent fluctuation exchange approximation. The calculated
magnetic response exhibits an anisotropic feature with largest momentum
span along the direction transverse to the momentum transfer $\mathbf{Q}_{\text{AFM}}$,
which gives rise to an elliptical image of the magnetic excitation.
The calculated momentum anisotropy of the magnetic spectrum agrees
with the INS measurements. An analysis on the orbital character of
the electronic structure associated with the deviation from perfect
nesting shows that the transversely lengthened short-range spin fluctuations
enhance intra-orbital, but inter-band, pair scattering processes that
play an important role to the formation of $s^{\pm}$-wave superconductivity
in this system. Therefore, this anisotropic momentum structure of
the magnetic fluctuations favors the development of the SC phase in
the magnetic scenario for the iron-based superconductors.

Our detailed study on the resonance mode in the magnetic spectrum
shows that the dispersion of the mode is also anisotropic with larger
broadening along the transverse than the longitudinal direction. Meanwhile,
the mode propagates upwards with increasing energy in the case of
nearly isotropic $s^{\pm}$-state and commensurate spin susceptibility,
but vanishes above the particle-hole threshold.

As the feedback from the spin excitations on fermions, the spectral
function exhibits the peak-dip-hump feature, which serves as one of
the interpretations of the ARPES observation of the kink feature.
\begin{acknowledgments}
We would like to thank A. V. Chubukov, I. Eremin, R. M. Fernandes,
and R. J. McQueeney for helpful discussions and Charles Zaruba for
technical help. This work was supported by the U.S. Department of
Energy, Office of Basic Energy Sciences, DMSE. Ames Laboratory is
operated for the U.S. DOE by Iowa State University under Contract
No. DE-AC02-07CH11358. \end{acknowledgments}

\end{document}